# High-quality Two-Dimensional Electron Gas in Undoped InSb Quantum Wells

Zijin Lei[1]*, Erik Cheah[1], Km Rubi[2], Maurice E. Bal[2], Christoph Adam[1], Rüdiger Schott[1], Uli Zeitler[2], Werner Wegscheider[1], Thomas Ihn[1], and Klaus Ensslin[1]

*Corresponding author: zilei@phys.ethz.ch

1. Solid State Physics Laboratory, ETH Zurich, CH-8093 Zurich, Switzerland

2. High Field Magnet Laboratory (HFML-EMFL), Radboud University, 6525 ED Nijmegen, the Netherlands

## Abstract

We report on transport experiments through high-mobility gate-tunable undoped InSb quantum wells. Due to the elimination of any Si modulation doping, the gate-defined two-dimensional electron gases in the quantum wells display a significantly increased mobility of 260,000 cm$^2$/Vs at a rather low density of $2.4\times10^{11}$ cm$^{-2}$. Using magneto-transport experiments, we characterize spin-orbit interactions by measuring weak antilocalization. Furthermore, by measuring Shubnikov–de Haas oscillations in tilted magnetic fields, we find that the *g*-factor agrees with ***k·p*** theory calculations at low magnetic fields but grows with spin polarization and carrier density at high magnetic fields. Additionally, signatures of Ising quantum Hall ferromagnetism are found at filling factor ν = 2 for tilt angles where the Landau level energy equals the Zeeman energy. Despite the high mobility, the undoped InSb quantum wells exhibit no fractional quantum Hall effect up to magnetic fields of 25 T.

## I. Introduction

InSb is a narrow-gap III-V compound known for its light effective mass, large *g*-factor in bulk material, and strong spin-orbit interactions (SOIs) [1-5]. These unique properties are interesting for potential applications such as high-frequency electronics [1], optoelectronics [6], and spintronics [7]. Recently, InSb, together with InAs, has attracted attention as a potential platform for topological quantum information processing [8]. As introduced in a series of both theoretical and experimental works, a topological nontrivial phase can be achieved in a semiconductor-superconductor hybridized device. It needs a sophisticated control of Zeeman energy, chemical potential, Rashba SOI, and the phase of supercurrent if the system is two-dimensional (2D) to form a pair of Majorana zero modes on sample terminals [9, 10, 11, 12]. Currently, experimental works on Majorana physics are primarily focused on InAs and InSb nanowires and InAs quantum wells (QWs), while the progress in InSb QWs is still hampered, and related publications are rare [11]. Although InSb QWs have been successfully grown using molecular beam epitaxy (MBE) methods [13], the technique to obtain high-quality two-dimensional electron gases (2DEGs) in InSb QWs is still not as mature as that in InAs QWs. For instance, the mobility of InAs QWs has reached values as high as ~2,400,000 $cm^2/Vs$, while this value is still around 300,000 $cm^2/Vs$ in InSb QWs [14-16]. In addition, the carrier density in InSb QWs is not as stable as those in InAs and GaAs. A pinched-off channel always becomes conductive again after minutes when keeping gate voltages constant. Due to this shortcoming, nanoconstrictions based on InSb QWs still pose a number of challenges [17, 18]. Recently, Kulesh *et al*. [19] presented a purely gate-defined quantum dot in an InSb QW. The instability of carrier density with time is best reduced by eliminating Si-doping during the growth of the heterostructures. Then, the carriers in these undoped InSb QWs are induced through an electric field applied by a global top gate. Though stable Coulomb blockade effects were achieved in their device [19], undoped InSb QWs need further optimization. As illustrated in the work of undoped GaAs/AlGaAs heterostructures, the elimination of modulation doping minimizes remote ionized dopant scattering [20, 21]. With the previous work in various III-V compound heterostructures as references, we expect that 2DEGs with both high mobility and stable density can be induced in undoped InSb QWs.

In this paper, we introduce undoped InSb QWs as a platform for magneto-transport experiments. Due to the elimination of modulation doping, gate-defined 2DEGs display a significantly increased mobility compared to modulation-doped devices at low carrier densities less than $2.4 \times 10^{11}$ cm$^{-2}$. Second, we manage to probe the tunable SOIs of the 2DEGs with the measurement of weak antilocalization (WAL). Furthermore, by coincidence measurements of the Shubnikov–de Haas (SdH) oscillations in tilted magnetic fields, we find that the *g*-factor agrees with ***k·p*** theory calculations in low magnetic fields but grows with spin polarization and carrier density in high magnetic fields. Additionally, these 2DEGs show signatures of Ising quantum Hall ferromagnetism when the Zeeman energy equals the Landau energy in tilted magnetic fields. Finally, fractional quantum Hall effects are still absent in high magnetic fields even in these high mobility devices.

## II. Sample preparation and measurement methods

The InSb QW sample we investigate here is grown on a (100) GaAs substrate by molecular beam epitaxy (MBE). The growth process is based on a previous publication by Lehner *et al*. [13]. First, a specialized interfacial misfit transition to a GaSb buffer and an interlayer InAlSb buffer is employed to overcome the lattice mismatch between GaAs and InSb. The total thickness of the buffer system amounts to roughly 3 $\mu$m. Then, the 21 nm-thin InSb QW is surrounded by $In_{0.9}Al_{0.1}Sb$ confinement barriers. On top of the QW, the thickness of the $In_{0.9}Al_{0.1}Sb$ barrier is 50 nm. There is no Si modulation doping performed during growth.

In contrast to the doped structure, the micro-fabrication of undoped InSb needs more optimization and additional steps and is presented in the following. First, a standard $400\mu m \times 200\mu m$ Hall bar is defined using chemical etching with an etching depth of ~ 80 nm, deeper than the QW. Similar to the undoped GaAs/AlGaAs heterostructures, ohmic contacts are more challenging to achieve in undoped InSb QWs [20, 21]. Here, layers of metals are evaporated after an Ar milling step to provide Ohmic contacts. Because there are no carriers without an electric field applied, we need to anneal the sample for the contact material to diffuse into the QWs both in the growth and the lateral direction. Therefore, the chance of avoiding an insulating gap between the alloyed contacts and 2DEGs is maximized. Ge/Ni/Au as contact metals are adopted because this alloy diffuses into III-V semiconductors even at relatively low temperatures. We also recommend a shallow (~10 nm) etch on the areas where contacts locate to remove the oxidized layer on the surface and decrease the distance of metal diffusion to the QWs. In the next step, the sample is coated with a 40-nm-thick aluminum oxide (ALO) dielectric layer using atomic layer deposition (ALD) at a temperature of 150℃ for 1.5 hours. These ALD parameters are optimized considering metal diffusion, dielectric layer quality, and preserving the mobility of 2DEGs. On the one hand, a longer heating time or higher temperature will reduce the mobility of the 2DEG, though the ALO layer will have better quality. On the other hand, the metal will not diffuse sufficiently if the heating is not enough. Finally, a Ti/Au top gate covering the Hall bar and the contacts is deposited by electron-beam evaporation. With this method, we successfully made 11 working devices from 3 nominally identical wafers. For all working samples, the contacts are measured to be ohmic, and the 2-terminal resistances of samples are within 10 kΩ when the carrier densities are maximum.

The transport measurements are accomplished using standard low-frequency lock-in techniques in a cryostat with a base temperature of 1.7 K. First, with the top gates grounded, the QWs are completely insulating. We verify that a high-quality single-channel InSb 2DEGs is formed when a positive top gate voltage $V_{TG}$ is applied. Figure 1 (a) shows the density $n$ determined via the classical Hall effect in small magnetic fields and the corresponding Drude mobility $\mu$ of the 2DEG as a function of $V_{TG}$. This measurement is performed on a Hall bar oriented along the crystal direction [1 -1 0]. With the increase of $V_{TG}$, $n$ increases up to $2.4 \times 10^{11}$ cm$^{-2}$, and the corresponding $\mu$ increases to 260,000 cm$^2$/Vs. The density $n$ here is limited because of the breakthrough of the ALO dielectric layer when a large electric field is applied. Compared with Si-doped InSb of our previous work [16], the mobility increases by ~60 % for equivalent density. This mobility is more than twice as high as in the undoped wafer used in the work by Kulesh *et al*. [19] at the same density, and it is comparable with the previous doped wafer in Ref. [2], where, nevertheless, the carrier density cannot easily be tuned. In our samples, the gate capacitance $C_{\text{Hall}}$ is estimated to be 0.47 mF/m$^2$ from the linear part of the $n$-$V_{\text{TG}}$ line, which is only about 40% of the value calculated from the geometry of the layer structure through a parallel-plate capacitor model $C_{\text{Geo}}$. We found this mismatch between expected and measured capacitance consistently for a number of samples and substrates, doped and undoped structures. This mismatch has been reported previously by both our group and another group independently [16, 19]. Figure 1 (b) shows the $n$ - $\mu$ traces of both the Hall bars along crystal orientations [1 -1 0] and [1 1 0], respectively. The data for the [1 -1 0] direction is the same as in Fig.1 (a). We find an anisotropy of the mobility in the two perpendicular crystal orientations, probably related to an anisotropic growth. For comparison, we also tried to grow undoped InSb QWs on undoped GaSb substrates with the same growth sequence. Because there is much less lattice mismatch between GaSb and InSb, the growth of the buffer layer is not necessary and therefore not implemented. However, we did not manage to produce conducting Hall bars with the same micro-fabrication process. Interestingly, a 2DEG with mobility $\mu$ = 330,000 cm$^2$/Vs at density $n$ = 2.4×10$^{11}$ cm$^{-2}$ is induced by $V_{\text{TG}}$ in QWs grown on GaSb substrate when a tiny amount of modulation Si-doping is added. Here, the doping dose is so low that a 2DEG cannot be induced when the top gate is grounded. However, the channel of the QW is "switched-on" with a positive $V_{TG}$ of less than 1 V applied. The quality in terms of mobility of this InSb grown on GaSb substrate with tiny Si-doping is even higher. Nevertheless, we find that the ohmic contacts on InSb QWs with GaSb substrate are much worse than those on GaAs substrates because the contacts do not work in magnetic fields higher than 3 T.

## III. Weak antilocalization measurement

The high-quality 2DEGs in undoped InSb QWs also display tunable SOI, which is characterized by WAL measurements. Figure 2 (a) presents the WAL effects at temperatures $T$ varying from 1.7 K to 9 K for $n = 2.1 \times 10^{11}$ cm$^{-2}$. To increase the signal-to-noise ratio, each trace presented here is an average of more than 5 consecutive measurements. We find the conductivity peak of WAL at zero magnetic field by converting the longitudinal resistance and Hall resistance to longitudinal conductivity $\sigma_{xx}$ and subtracting a polynomial background. The height of the peak decreases with increasing $T$. The precise measurement of $\sigma_{xx}(B)$ enables us to extract coherence and SOI lengths using the Hikami-Larkin-Nagaoka (HLN) expression [22, 23]. Similar methods are also adopted in papers where WAL effects are measured in doped InSb QWs [11] and InSb nanosheets [24]. The conductivity correction $\Delta\sigma_{xx}(B)$ of WALs reads:

$$\Delta\sigma_{xx}(B) = \frac{e^2}{2\pi^2\hbar}\left(\psi\left(\frac{1}{2} + \frac{H_\phi}{B} + \frac{H_{SO}}{B}\right) + \frac{1}{2}\psi\left(\frac{1}{2} + \frac{H_\phi}{B} + \frac{2H_{SO}}{B}\right) - \frac{1}{2}\psi\left(\frac{1}{2} + \frac{H_\phi}{B}\right) - \ln\left(\frac{H_\phi + H_{SO}}{B}\right) - \frac{1}{2}\ln\left(\frac{H_\phi + 2H_{SO}}{B}\right) + \frac{1}{2}\ln\left(\frac{H_\phi}{B}\right)\right).$$

Here, $H_\phi$ and $H_{SO}$ are phase coherence field and spin-orbit field, respectively. These two fitting parameters can be converted to phase coherence length $l_\phi$ and spin-orbit length $l_{SO}$. Using $l_\phi = \sqrt{\frac{\hbar}{4eH_\phi}}$, we present $l_\phi$ as a function of $T$ in Fig. 2(b) when $n$ = 1.7×10$^{11}$ cm$^{-2}$ and 2.1×10$^{11}$ cm$^{-2}$. The dashed line here is a fitting of the power-law $l_\phi \propto T^{-\alpha}$. In both sets of measured data, $\alpha$ is very close to 1. This implies that our system is in the standard 2D diffusive regime. A similar $l_\phi \propto 1/T$ relation is observed in other 2DEG systems [25-27]. Additionally, a 3-parameter model by Iordanskii, Lyanda-Geller, and Pikus [23], which in principle allows extracting Rashba and Dresselhaus contributions separately, does not give additional information for the range of data considered here.

The SOI of the InSb 2DEG can be tuned by $n$ with $V_{TG}$, which is probed with WAL measurements as well. Figure 3 (a) shows the WAL with $n$ varying from 1.67×10$^{11}$ to 2.2×10$^{11}$ cm$^{-2}$ at 1.7 K. Here, we utilize the HLN expression again to analyze the SOI as a function of $n$. Figure 3 (b) presents the splitting $\Delta_{SO}$ vs. $n$ at the Fermi level due to SOI. Here $\Delta_{SO}$ is obtained with $\Delta_{SO} = \sqrt{\frac{2\hbar^2}{\tau_D \tau_{SO}}}$. Note that the spin-orbit time $\tau_{SO}$ is calculated with the formula $\tau_{SO} = \frac{\hbar}{4eDH_{SO}}$, where $D$ is the diffusion constant in 2D. With the increase of $n$ in the range of the measurements, $\Delta_{SO}$ increases from 0.5 to 0.74 meV. The values obtained here agree with the previous publication by Ke *et al.* [11], where their $\Delta_{SO}$ is slightly larger due to a higher $n$. Then, we normalize $\Delta_{SO}$ to get the coefficient $\alpha_{SO}$ of the SOI with $\alpha_{SO} = \frac{\Delta_{SO}}{2k_F}$, where $k_F$ is the Fermi wave vector. As shown in the inset of Fig. 3(b), $\alpha_{SO}$ also increases when $n$ is higher. Because InSb is a material with both strong Rashba and Dresselhaus SOIs, we expect that the contribution of Rashba and Dresselhaus SOIs are comparable. Additionally, since the Dresselhaus coefficient $\beta_D$ is an intrinsic material parameter, the increase of $\alpha_{SO}$ is likely due to a larger Rashba contribution from a higher electric field in the high-density regime. Furthermore, as presented in Fig. 3 (c), $l_\phi$ increases when $n$ is higher, which means that the decoherence is reduced with higher density and mobility.

## IV. Scattering mechanism analysis

Next, we further study these high-quality InSb 2DEGs with magneto-transport measurements in magnetic fields up to 8 T. As shown in Fig.1 (b) before, in our 2DEGs, $\mu$ is proportional to $n$. This indicates that long-range remote ionized impurity scattering and short-range background impurity scattering are comparable [28-30]. More insight into the scattering mechanisms can be obtained from analyzing SdH oscillations with the Ando formula [27]. Figure 4 (a) presents the effective mass of InSb 2DEGs $m^*$ as a function of magnetic field $B$ when $n$ is $1.4 \times 10^{11}$ cm$^{-2}$ and $2.1 \times 10^{11}$ cm$^{-2}$. The method to measure $m^*$ is similar to our previous publications [16, 31]. In the range with small amplitude SdH oscillations compared to the magneto-resistance background, $m^*$ has a roughly constant value of 0.015 $m_e$ in both low- and high-density regimes, where $m_e$ is the free electron mass. This shows that the non-parabolicity of the conduction band is still negligible within the range of $n$ in which we are interested. This value agrees with the $\boldsymbol{k} \cdot \boldsymbol{p}$ calculations and our previous measurement in QWs with the same width [16]. With the obtained $m^*$ plugged into the Ando formula, the quantum lifetime of Landau levels $\tau_q$ is extracted in the following. Figure 4 (b) shows $|\Delta\rho_{xx}|/(\bar{\rho}_{xx} f(B,T))$ vs. $1/B$ with $n = 1.4 \times 10^{11}$ cm$^{-2}$ at 2.1 K. Here $\Delta\rho_{xx}$ is the oscillating part of the magneto-

resistance obtained by subtracting a polynomial background $\bar{\rho}_{xx}$, and $f(B,T) = \frac{\frac{2\pi^2 k_B T}{\hbar \omega_c}}{\sinh\left(\frac{2\pi^2 k_B T}{\hbar \omega_c}\right)}$. Then, $\tau_q$ is extracted to be 0.067 ps using a linear fit. This value is much larger than the Drude mean free time $\tau_D$, which is 1.26 ps calculated from the Drude mobility when $B = 0$ with the same $n$. The large ratio of $\tau_D$ vs. $\tau_q$ implies a significant contribution of long-range scattering [32, 33].

Knowing $\tau_q$, we can estimate the g-factor of the 2DEGs from an activation energy measurement. Figure 4 (c) plots the longitudinal conductance $\sigma_{xx}$ of filling factor $\nu = 1$ and $\nu = 2$ as a function of $T$ with $n = 1.4 \times 10^{11}$ cm$^{-2}$. The $\sigma_{xx}$-$T$ diagrams follow the expression $\sigma_{xx} \propto \exp(-E_{A,1(2)}/2k_B T)$, where $E_{A,1(2)}$ is the activation energy of filling factor $\nu = 1(2)$. From the linear fit in Fig. 4 (c), $E_{A,1}$ and $E_{A,2}$ are found to be 12.78 meV and 8.41 meV, respectively. Without considering the Landau level broadening due to scattering, we can write $g^* \mu_B B \approx E_{A,1}$ at filling factor $\nu = 1$. Therefore, the g-factor at filling factor $\nu = 1$ is $g^* \sim 36$. Though this is a lower limit of $g^*$, it is already close to the value calculated from the $\boldsymbol{k} \cdot \boldsymbol{p}$ model [16]. A better estimation of the g-factor through the activation energy is the following. As shown in Fig. 4(d) schematically, with the consideration of a Landau level broadening $\Gamma$, the activation energies $E_{A,1}$ and $E_{A,2}$ then read $E_{A,1} = g^* \mu_B B - \Gamma$ and $E_{A,2} = \hbar \omega_c - g^* \mu_B B - \Gamma$, respectively. Here $\hbar \omega_c$ is the Landau level energy. Assuming $\Gamma = \sqrt{\frac{1}{2\pi} \hbar \omega_c \frac{\hbar}{\tau_q}}$ as suggested in Ref. [27], $g^*$ for $\nu = 1$ and $\nu = 2$ are calculated to be about 64 and 46, respectively. These values are larger than the calculation from the single-particle picture, where $g^* \sim 39$ with the same width of QWs. More about the larger *g*-factor will be discussed in the following section.

## V.  Coincidence measurement

Next, we present our results of the coincidence measurement. This method has been introduced in previous publications [16, 34-36]. The Landau level energy is proportional to the perpendicular magnetic field $B_\perp$, while the Zeeman energy is proportional to the total magnetic field $B_{tot}$. The ratio between these two energies can be changed continuously by changing the angle $\theta$ between the direction of sample normal and the total magnetic field $B_{tot}$ (inset of Fig. 5(a)). Therefore, $B_\perp$, the projection of the magnetic field along the sample normal, is $B_\perp = B_{tot}\cos(\theta)$. Here we introduce a parameter $r = g^*\mu_B B_{tot}/\hbar\omega_c$. Hence, we obtain $r\cos(\theta) = g^*m^*/2m_e$. The number of *r* can be obtained by monitoring the behavior of local maxima and minima of $\rho_{xx}$ at different tilt angles $\theta$. In a single-particle picture, for instance, at *r* = 1, i.e., when the Zeeman energy equals the Landau energy, the minima of SdH traces of longitudinal resistivity $\rho_{xx}$ will only occur at odd filling factors, such as $\nu$ = 3, 5, and 7. In contrast, at *r* = 2, the minima of $\rho_{xx}$ occur at even filling factors, such as $\nu$ = 4, 6, and 8. Additionally, the minima of $\rho_{xx}$ for $\nu$ = 1 and 2 remain the minima when *r* = 2 because the gaps are always open.

Figure 5 (a) depicts the dependence of $\rho_{xx}$ on $B_\perp$ with a continuous change of $\theta$ at 0.5 K when $n = 2.4\times10^{11}$ cm$^2$. The value of $\theta$ is calibrated with the slope of the Hall trace at low magnetic field with high accuracy. For a better determination of *r* = 1 and 2, we extract the $\rho_{xx}$ with even ($\nu$ = 4, 6, and 8) and odd filling factors ($\nu$ = 5 and 7) and plot them in Fig. 5 (b) and (c) as functions of $\theta$. For each even filling factor, the coincidence of $r = 1$ is found by picking the $\theta$ at which $\rho_{xx}$ reaches its maximum. Similarly, the coincidence of $r = 2$ of each odd filling factor is found where $\rho_{xx}$ has a maximum. The error bar of each coincidence angle is determined by the step of $\theta$ between the coincidence and the nearest data points around it.

Moreover, the gap at $\nu = 2$ stays open at low temperature with the coincidence condition $r = 1$. This is different from the situations of $\nu = 4, 6,$ and 8. For $\nu = 2$ $\rho_{xx}$ stays zero for all tilt angles. Nevertheless, the minimum of $\rho_{xx}$ at $\nu = 2$ starts to increase above zero when $T > 1$ K. Through reproducing the SdH measurements at 1.5 and 4.2 K with a fine-tuning of $\theta$, we determine that the gap at $\nu = 2$ reaches its minimum when $\theta = 66.22°$, where the uncertainty of $\theta$ is less than 0.5°. This also means that the coincidence condition of $r = 1$ is most precisely achieved at this angle. Figure 6 depicts the $T$-dependence of the longitudinal resistivity $\rho_{xx}$ and transversal resistivity $\rho_{xy}$ with the coincidence condition $r = 1$ for filling factor $\nu = 2$. At filling factor $\nu = 2$, $\rho_{xx}$ increases with increasing $T$ from 0.35 K to 8 K. Meanwhile, the Hall plateau at $\nu = 2$ becomes narrower and approaches the classical Hall effect at 8 K, though plateaus for higher filling factors, such as $\nu = 3$ and 5, are still pronounced. This is different from the previous paper by Chokomakoua et al., where an InSb QW in van der Pauw geometry was studied [37]. There, the Hall plateau of $\nu = 2$ was not well defined at low temperatures. However, our results are similar to the work by Koch *et al.*, where the Landau level anti-crossing due to a magnetic instability is found in GaInAs/InP heterostructures [38].

In addition to a nonvanishing gap at filling factor $\nu = 2$ with $r = 1$, a resistivity peak appears at intermediate temperature but vanishes at high temperature in $\rho_{xx} - B_\perp$ traces. This agrees with the phenomenon of quantum Hall Ising ferromagnetism [39, 40]. The inset in the upper panel of Fig. 6 depicts this process. The appearance and vanishing of the resistivity spikes can be quantitatively depicted using Curie temperature $T_C$. Here we adopt the methods introduced by De Poortere *et al.* to estimate $T_C$ [40]. The heights $\Delta\rho$ of the resistivity peaks at different temperatures are obtained by subtracting a polynomial magneto-resistance background around $\nu = 2$ (black dashed line in the inset of the upper panel of Fig. 6 (a)) from the measured $\rho_{xx}$. Plots of $\Delta\rho$ vs. $T$ are shown in the upper inset of the lower panel of Fig. 6. When $T = T_M = 3.15$ K, $\Delta\rho$ reaches its maximum. Meanwhile, the width of the resistivity spike starts to increase when $T \geq T_M = 3.15$ K as well. Therefore, we estimate the Curie temperature $T_C$ to be around a value $T_C \cong T_M \sim 3$ K. Though the phenomenon of Ising ferromagnetism is less pronounced at lower density, we still managed to estimate $T_C$ to be $\sim 3$ K when $n = 1.7 \times 10^{11}$ cm$^2$. In addition, the *T*-dependence of $\rho_{xx}$ at $\nu = 2$ meets the thermal activation [41] model numerically within the accuracy of our measurement. Similar to Fig. 4 (c), we analyze the $\sigma_{xx}$-$T$ diagram with $\sigma_{xx} \propto \exp(-E_A/2k_BT)$ and obtain $E_A = 0.44$ meV (lower inset of the lower panel in Fig. 6). We reproduced this measurement in a sample with the same quality when $n = 1.7 \times 10^{11}$ cm$^{-2}$, finding that $E_A = 0.18$ meV.

With the coincidence condition of $r = 1$ at $\nu = 2$ included, we investigate the spin-polarization dependence of the spin susceptibility $\chi$. From the data presented in Fig. 5, the coincidence angles with different $r$ and $\nu$ are extracted. Using $r\cos(\theta) = g^*m^*/2m_e$, $g^*m^*$ is obtained. Because $\chi = \frac{g^*m^*}{2\pi\hbar^2}$ in our 2D system, we use $g^*m^*$ to describe the spin susceptibility in this work. Figure 7 (a) shows the traces of $g^*m^*$ as a function of spin polarization $P$ when $n$ is 2.4×10¹¹ cm² (blue) and 1.4×10¹¹ cm² (red), respectively. As introduced by Zhu *et al.* [42], $P$ can be calculated easily through $P = r/\nu$. In both traces presented, $g^*m^*$ grows with $P$. For a better illustration, we present $g^*$ along the right axis, assuming that $m^*$, measured to be 0.015 $m_e$ from SdH oscillations, is also independent of $P$. For low magnetic fields, $g^*$ is ~ 40 in both cases, which agrees with the calculation within a $\boldsymbol{k}\cdot\boldsymbol{p}$ model. Nevertheless, when $P = 0.5$, $g^*$ increases significantly up to ~ 53 and ~ 46 with $n = 2.4\times10^{11}$ cm² and 1.4×10¹¹ cm², respectively. This $g^*$- $n$ relation agrees with the publications by Yang *et al.* [36] and Nedniyom *et al.* [43]

Furthermore, the spin susceptibility also grows with $n$. We summarize all the $g^*m^*$ measured with $r = 1$ and $\nu=2$ from the same three samples and plot them as a function of $n$ in Fig. 7 (b). Assuming that $m^*$ is constant at 0.015 $m_e$, $g^*$ is calculated and plotted too. With the increase of $n$ from $1.4\times10^{11}$ cm⁻² to $2.4\times10^{11}$ cm⁻², $g^*m^*$ increases by ~15%. This tendency is opposite to the results from dilute 2DEGs in GaAs, where $g^*m^*$ decreases with the increase of $n$ [42]. In GaAs 2DEGs, the interaction parameter $r_s$, defined as the ratio between Coulomb energy and Fermi energy, is smaller when $n$ is higher. Nevertheless, due to the light effective mass and the high density in InSb 2DEGs, $r_s$ is only between 0.17 and 0.23 in our experiment. This value is not only much smaller than those of low-density 2DEG in GaAs and AlAs, but the variation is also limited in our experiment. Therefore, the contribution of electron-electron interaction to the $n$ dependence of the spin susceptibility is not dominant. Instead of electron-electron interactions, this $g^*m^*$-$n$ diagram is more likely to be influenced by the SOI. As we verified through WAL measurement before, the SOI is stronger with increasing $n$. Effectively, this extra contribution is probed as an enlarged Zeeman energy through the coincidence measurement, leading to a larger $g$-factor. Similar phenomena have been observed in previous work in an InAs 2DEG [44], another narrow band-gap material with light effective mass and strong SOI, where the anti-crossing of Landau levels in titled fields was also found.

## VI. Magnetotransport when $\nu < 1$

Finally, motivated by the high mobility of our 2DEGs, we probe the behavior of the 2DEGs in even higher perpendicular magnetic fields. Figure 8 depicts one of the magneto-resistance measurements of our sample at 0.7 K with $n = 2.4 \times 10^{11}$ cm$^{-2}$. In $B$ fields where $\nu < 1$, the Ohmic contacts are still working properly, indicating metallic behavior of the sample [45]. However, despite the high quality of the sample, there are still no signatures related to fractional quantum Hall effects (FQHEs). The Hall resistance $\rho_{xy}$ increases with increasing $B$ when $\nu < 1$. However, it is only approaching but not reaching the classical limit (dashed black line). Meanwhile, there is no convincing local minimum in $\rho_{xx}$ or plateau in $\rho_{xy}$ when $B > 15$ T. The absence of any FQHE-related features shows that there is still room to increase the sample quality. We notice that the weak electron-electron interaction in InSb may not be advantageous to form fractional states. Furthermore, the quantum lifetime of InSb QWs is still not as long compared to systems where the FQHE has been observed, such as the pioneering works in GaAs, Si/Ge, and graphene [46-48]. The wide integer quantum Hall plateaus also indicate disorder hampering the formation of FQHEs in our devices. More optimized growth techniques are required to increase the chance in InSb to explore fractional states.

## VII.  Conclusion

In conclusion, we have presented transport experiments on gate tunable high-quality 2DEGs in undoped InSb QWs. With the elimination of Si modulation-doping, the mobility is significantly increased. Tunable SOIs were probed through WAL measurements in a 2D diffusive transport regime. Furthermore, using coincidence methods, we find that the *g*-factor grows with both the spin-polarization and carrier density. For filling factor $\nu = 2$, a signature of Ising quantum Hall ferromagnetic phase was observed in a tilted magnetic field and the Curie temperature was estimated. Finally, despite the high mobility, the undoped InSb QWs still did not exhibit fractional quantum Hall states up to magnetic fields of 25 T.

**Acknowledgment**

We thank Dr. F. K. de Vries, Dr. C. Reichl, and Mr. L. Ginzburg for fruitful discussions. We appreciate Mr. P. Märki and Mr. T. Bähler for their technical supports. We acknowledge the support of HFMLRU/NWO-I, member of the European Magnetic Field Laboratory (EMFL). We acknowledge funding from QuantERA. This paper was supported by the Swiss National Science Foundation through the National Center of Competence in Research (NCCR) Quantum Science and Technology.

**Figure 1.** (a) Data measured from the Hall bar along [1 -1 0] direction. The blue line shows the density $n$ as a function of the top gate voltage $V_{TG}$. The Drude mobility $\mu$ is presented vs. $V_{TG}$ with the red line. There is a mismatch between the capacitance obtained from Hall measurement $C_{Hall}$ and the parallel board model $C_{Geo}$. (b) The plots of $\mu$-$n$ when the Hall bar is along [1 -1 0] (red) and [1 1 0] (blue), respectively. Black dashed lines are a linear fit of $\mu$-$n$ with high density.

**Figure 2.** (a) The temperature dependence of weak antilocalization with $n = 2.1\times10^{11}$ cm$^{-2}$. The open circles are data points, and the solid black lines are fits using the HLN expression. The traces have a constant offset for a better presentation. (b) Phase coherence length $l_\phi$ vs. $T$ when $n = 1.7\times10^{11}$ cm$^{-2}$ (blue) and $2.1\times10^{11}$ cm$^{-2}$ (black) respectively. The dashed line is the fitting with the power-law $l_\phi \propto T^{-\alpha}$. The fitting shows that $\alpha$ = 0.95 and 0.94 when $n$ are $1.7\times10^{11}$ cm$^{-2}$ and $2.1\times10^{11}$ cm$^{-2}$, respectively.

**Figure 3.** (a) The density dependence of weak antilocalization with $T = 1.7$ K. The open circles are data points. Black solid lines are fits using the HLN expression. The traces are vertically offset for clarity. (b) The spin-orbit splitting at the Fermi level $\Delta_{SO}$ vs. density $n$. A tendency that $\Delta_{SO}$ increases with the increase of $n$ is observed. Inset: the spin-orbit coefficient $\alpha_{SO}$ as a function of $n$. $\alpha_{SO}$ also grows with $n$. (c) Phase coherence length $l_\phi$ vs. $n$. $l_\phi$ increases with higher $n$.

**Figure 4.** (a) Effective mass $m^*$ obtained from Ando formula fitting as a function of the magnetic field $B$. Tow sets of data with $n = 1.4 \times 10^{11}$ cm$^{-2}$ (black) and $2.4 \times 10^{11}$ cm$^{-2}$ (blue) are measured. Within the error bar from fitting, we determine $m^* = 0.015\ m_e$ as a constant value within the measurement range of our experiment. (b) The fitting to obtain quantum lifetime $\tau_q$ with $n = 1.4 \times 10^{11}$ cm$^{-2}$. The method is introduced in the main text. Here the black squares are the data points, and the dashed line is a linear fitting. (c) Temperature dependences of the longitudinal conductivity $\sigma_{xx}$ when $\nu = 2$ (red) and $\nu = 1$ (green), respectively. The activation energies $E_{A,2(1)}$ are obtained from the linear fits (dashed lines) presented in the figure. (d) A schematic diagram to depict the relationship between the Landau level broadening $\Gamma$, Zeeman energy $g^*\mu_B B$, Landau energy $\hbar\omega_c$, and the measured activation energy $E_{A,2}$ and $E_{A,1}$. In the figure, $E$ is the energy, and $D(E)$ is the density of states.

**Figure. 5** (a) The SdH oscillations measured for different tilt angles. The traces have a constant offset of 350 $\Omega$ for a better presentation. The definition of the tilted angle $\theta$ is presented in the inset. (b) The longitudinal resistance $\rho_{xx}$ with even filling factors $\nu = 4, 6,$ and $8$ vs. the tilted angle $\theta$. (b) The longitudinal resistance $\rho_{xx}$ with odd filling factors $\nu = 5$ and $7$ vs. the tilted angle $\theta$.

**Figure. 6** The temperature dependence of the longitudinal and transversal resistivity $\rho_{xx}$ and $\rho_{xy}$ when $\theta = 66.22°$. Here the coincidence $r = 1$ for $\nu = 2$ is achieved. Inset of the upper panel: the zoom-in of $\rho_{xx}$ - $B_\perp$ diagram around $\nu = 2$ with the same axis. The black dashed line is a polynomial background to show the resistivity peak. Upper inset of the lower panel: the peak height $\Delta\rho$ as a function of $T$. The Curie temperature $T_C$ is determined to be $T_C \sim 3.15$ K, where $\Delta\rho$ achieves its maximum. Here the stars are data, and the dashed line is guidance to eyes. Lower inset of the lower panel: analysis of $\sigma_{xx}$ vs. $T$ with the gap model when $\nu = 2$. The squares are the data and the dashed line is a linear fitting.

**Figure 7.** (a) $g^*m^*$ obtained from coincidence measurement vs. the spin polarization $P$ when $n = 2.4\times10^{11}$ cm$^{-2}$ (blue) and $1.4\times10^{11}$ cm$^{-2}$ (red), respectively. Plugging in $m^* = 0.015$ $m_e$, $g^*$ is obtained and shown in the axis of the right-hand side. (b) $g^*m^*$ and $g^*$ vs. $n$. Here we present the coincidence where $r = 1$ and $\nu = 2$. $m^*$ is still assumed to be 0.015 $m_e$.

**Figure 8.** $\rho_{xx}$ and $\rho_{xy}$ measurement of the InSb QW in a large magnetic field range at the temperatures of 0.7 K with $n = 2.4\times10^{11}$ cm$^{-2}$. The dashed line is an extrapolation of the Hall trace in the small magnetic field.

**Figure 1.** *(This is a one-column figure)*

a)
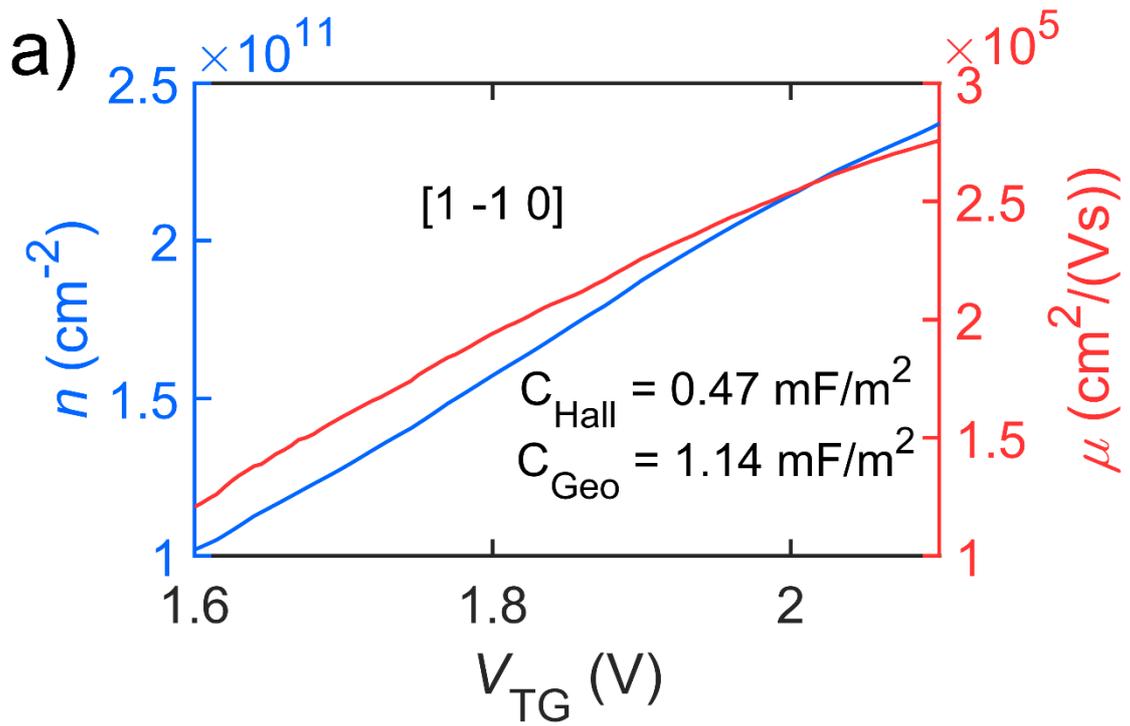

b)
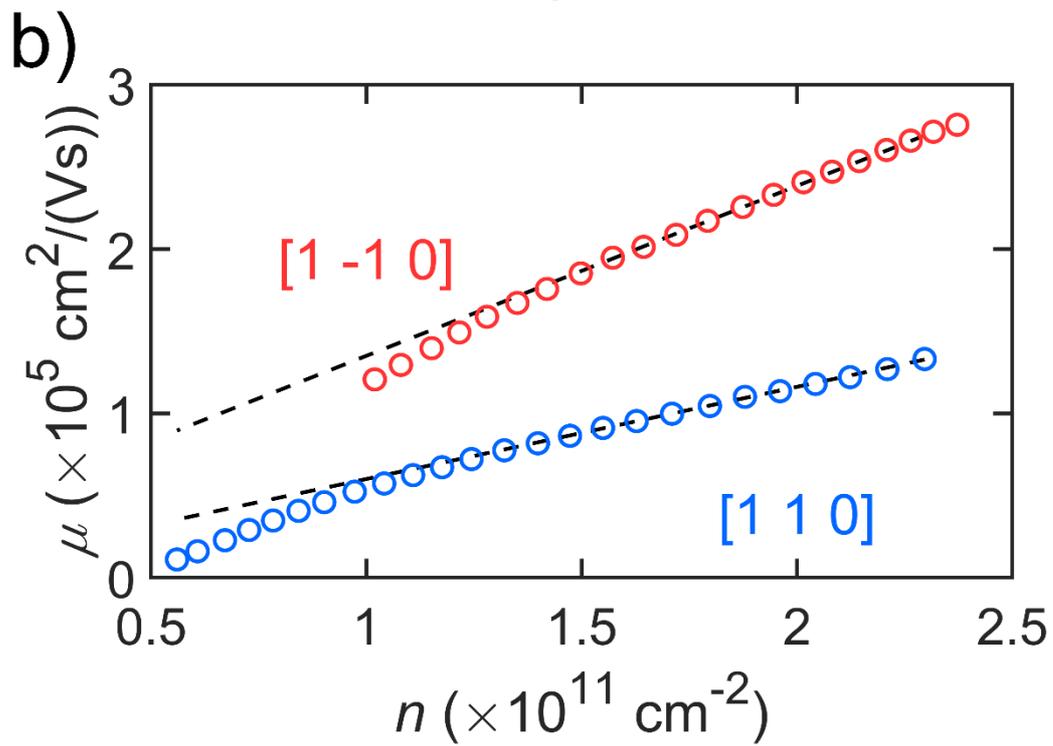

**Figure 2.** *(This is a one-column figure)*

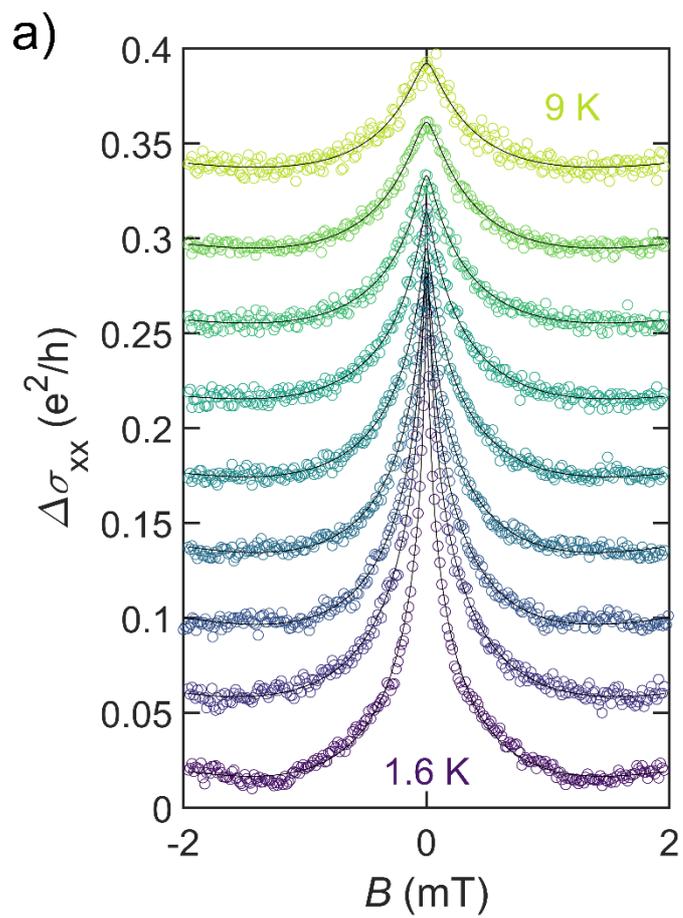

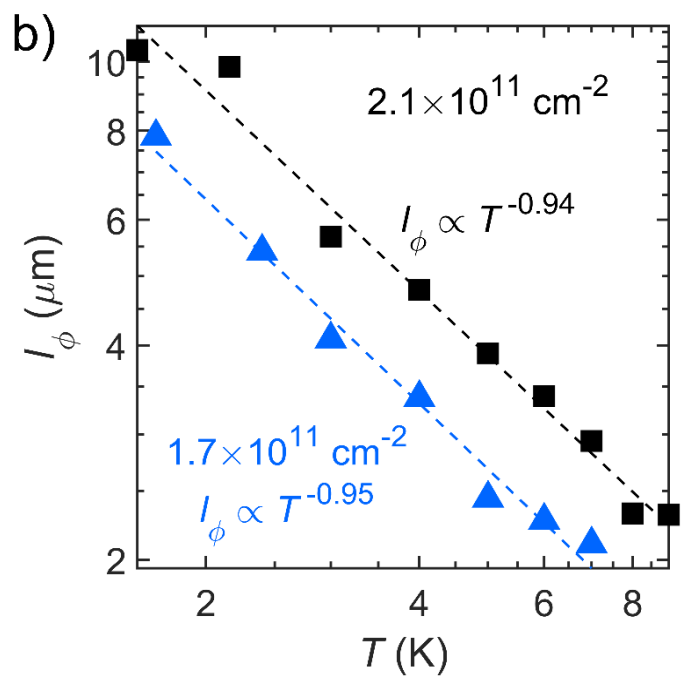

**Figure 3.** *(This is a one-column figure)*

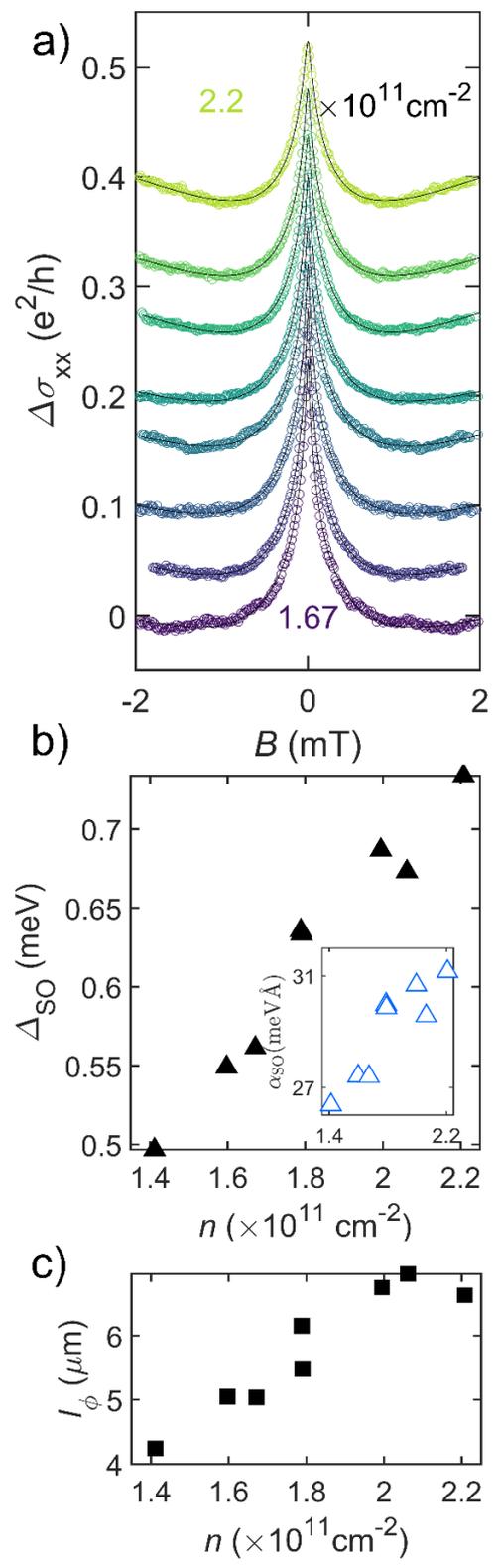

**Figure 4.** *(This is a two-column figure)*

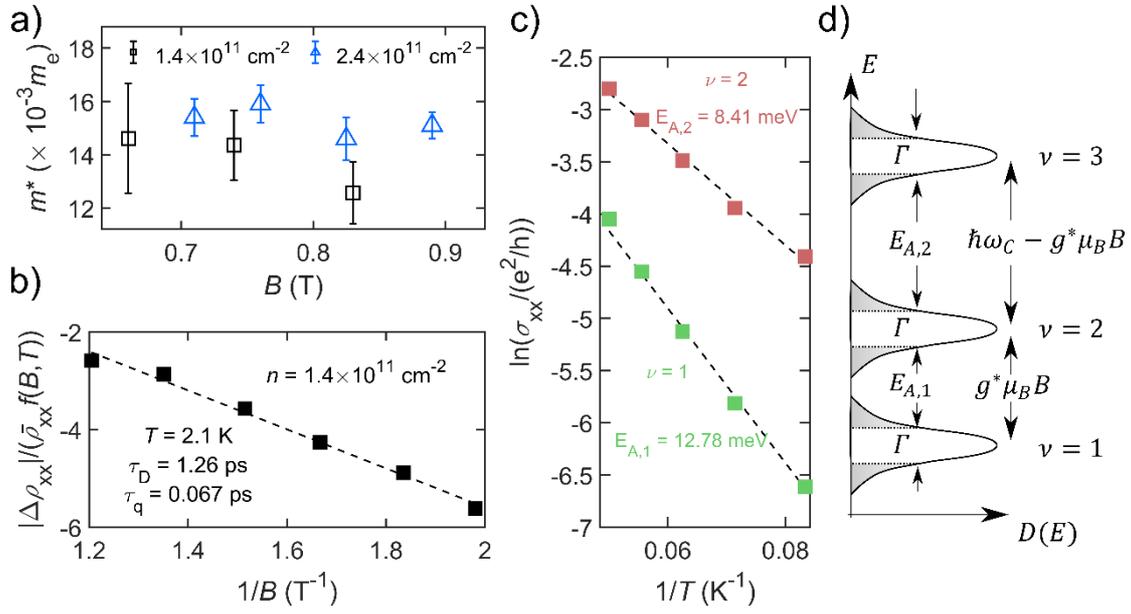

**Figure 5.** *(This is a two-column figure)*

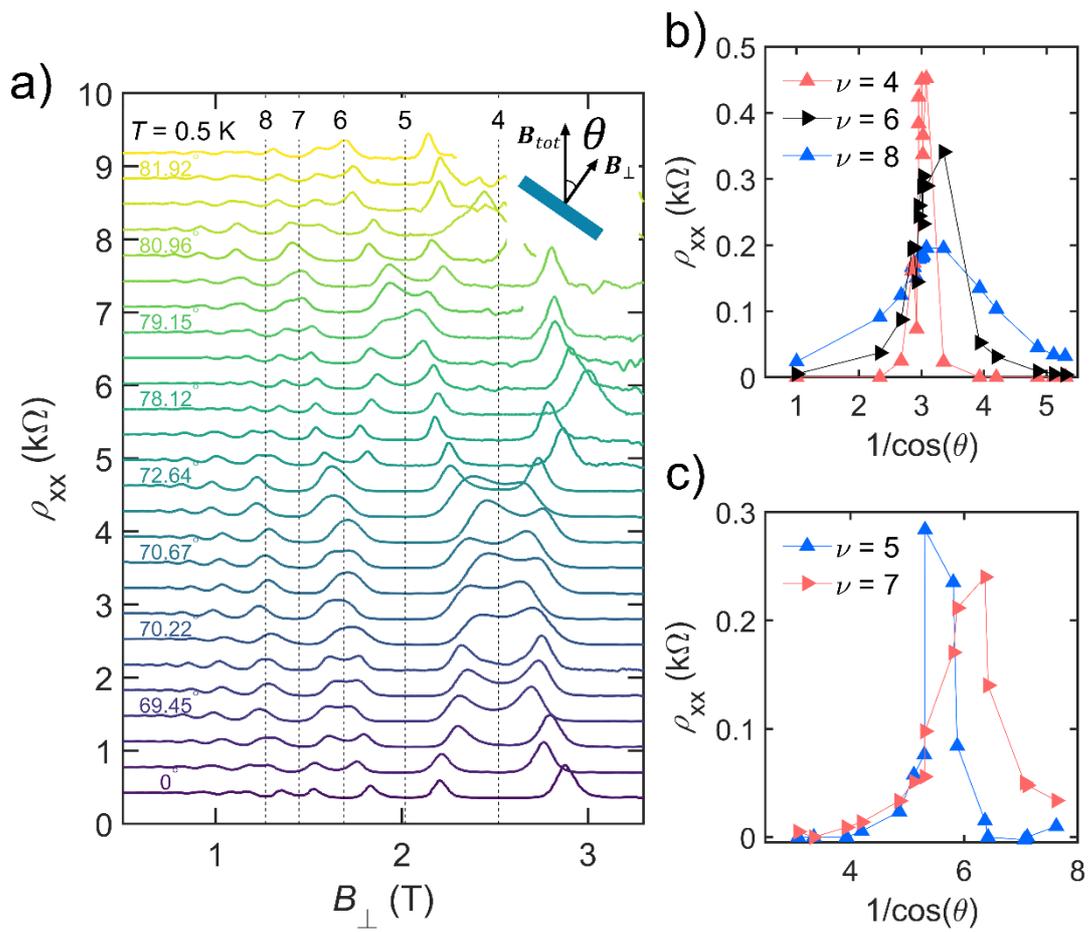

**Figure 6.** *(This is a one-column figure)*

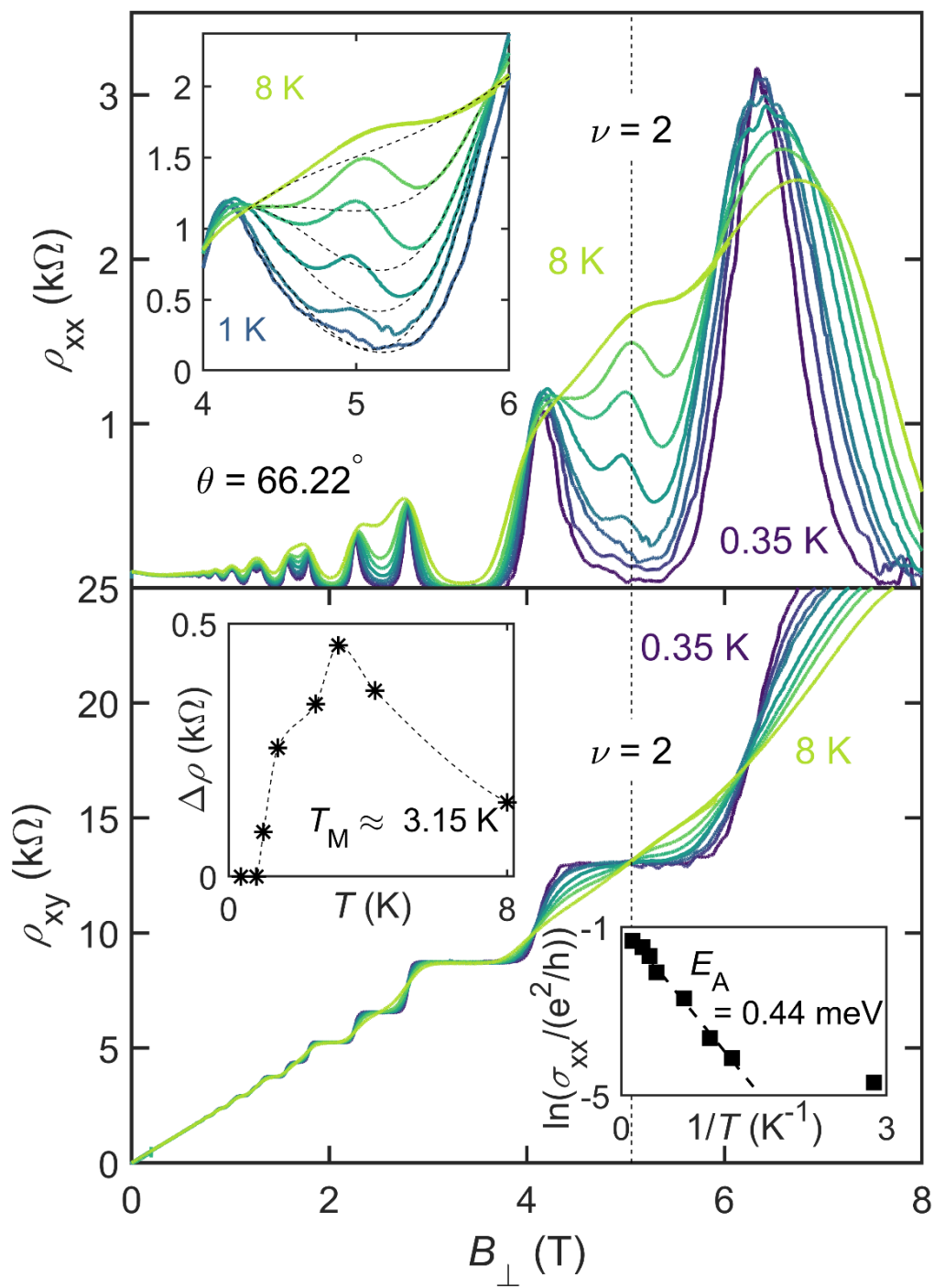

**Figure 7.** *(This is a one-column figure)*

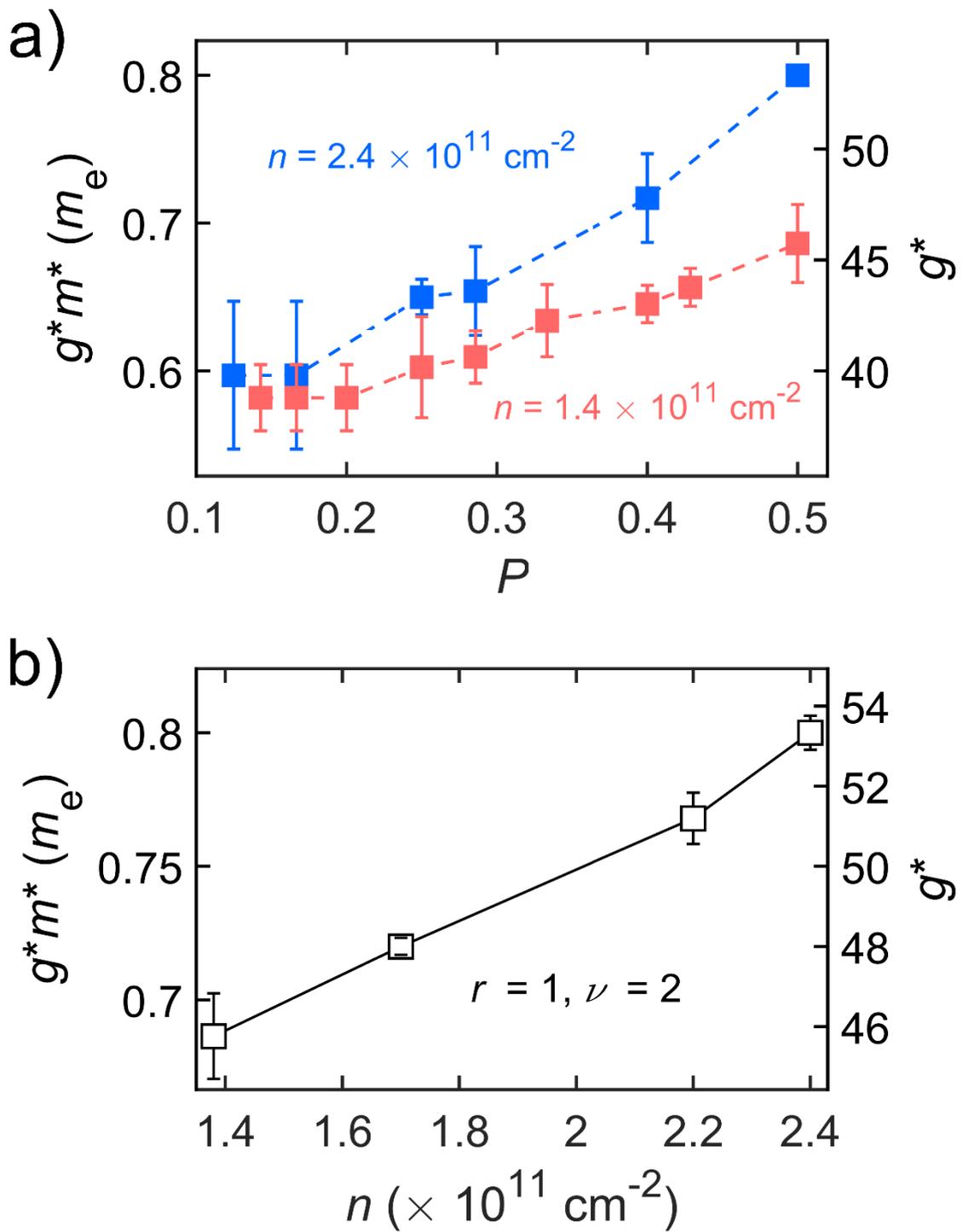

**Figure 8.** *(This is a one-column figure)*

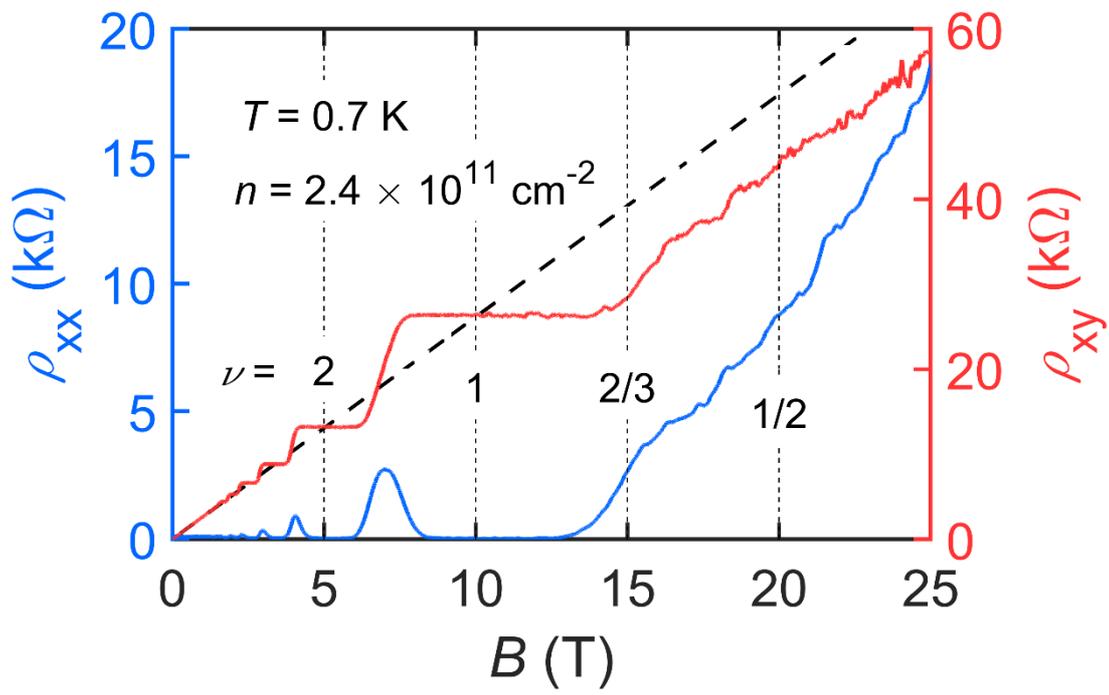